\documentclass[12pt]{article}
\usepackage{amsfonts,amssymb,amsmath}
\usepackage[dvips]{epsfig}
\begin{document}
\title{\bf Signature change by GUP }
\author{T. Ghaneh$^1$\thanks{Email: ghaneh@tabrizu.ac.ir}
\hspace{2mm},  \hspace{2mm}
 F. Darabi$^2$\thanks{Email: f.darabi@azaruniv.edu (Corresponding author)   }
 \hspace{2mm}, and \hspace{2mm}
 H. Motavalli$^1$\thanks{Email: motavalli@tabrizu.ac.ir } \\
\centerline{$^1$\small {\em Department of Theoretical Physics
and Astrophysics, University of Tabriz, 51666-16471, Tabriz,
Iran.}}\\
{$^2$\small {\em Department of Physics, Azarbaijan Shahid Madani University, 53714-161, Tabriz, Iran. }}}
\maketitle
\maketitle
\begin{abstract}
\noindent  
We revisit the issue of continuous signature transition from Euclidean to Lorentzian metrics in a cosmological model described by FRW metric minimally coupled with a self interacting massive scalar field. Then, using a noncommutative phase space of dynamical variables deformed by Generalized Uncertainty Principle (GUP) we show that the signature transition occurs even for a model described by FRW metric minimally coupled with a free massless scalar field accompanied by a cosmological constant. This indicates that the continuous signature transition might have been easily occurred at early universe just by a free massless scalar field, a cosmological constant and a noncommutative phase space deformed by GUP, without resorting to a massive scalar field having an {\it ad hoc} complicate potential. We also study the quantum cosmology of the model and obtain a solution of Wheeler-DeWitt equation which shows a good correspondence with the classical path.
\end{abstract}

\section{Introduction}

The idea of noncommuting coordinates firstly was proposed by Wigner
\cite{Wigner} and separately by Snyder \cite{Snyder}. This idea has been followed by Connes\cite{Connes} and Woronowicz \cite{Woronowicz} as noncommutative (NC) geometry, leading to a new formulation of quantum gravity through NC differential calculus \cite{Madore}. The link between NC geometry and string theory has also become evident by Seiberg and Witten \cite{Seiberg&Witten}, which resulted in NC field theories via the NC algebra based on the Moyal product \cite{Moyal}. Riemannian geometry of noncommutative surfaces has extensively been studied by Chaichian {\it et al} where they have developed a Riemannian geometry of noncommutative surfaces as a first step towards the construction of a consistent noncommutative gravitational theory \cite{Masud1}, which is relevant to the present paper. Possible effect of spacetime noncommutativity on primordial gravitational waves in inflationary cosmology has also been studied \cite{Cai}. Moreover, the fact that spacetime noncommutativity could suppress quantum fluctuations of matter fields, and dramatically constrain the random walking regime of the inflaton field at high energy scale is shown in \cite{Cai1}.

In recent years, the existence of a minimal observable length has been predicted by different aspects in merging gravity with quantum theory of fields
\cite{Gross&Mendle07,Gross,Amati&Ciafaloni&Veneziano,Kato,Konishi&Paffuti&Provero,Garay,Haro,Witten,Amelino&et.al.}. First, it was derived from string theory \cite{Amati&Ciafaloni&Veneziano,Witten,Veneziano}. In the spirit of perturbative string theory, this comes from the fact that strings can not probe distances smaller than the string size. This is the natural cut-off length at which the quantum effects of gravitation become considerable in comparison with the electroweak and strong interactions and the transparent smooth view of the very notion of the space-time becomes opaque. When the energy of a string reaches the Planck mass, the excitations of string may cause a nonzero extension \cite{Witten}. But creative calculations \cite{Maggiore65} show that this prediction is more reliable in quantum gravity and is not necessarily related to high energy or short distance behavior of the strings \cite{Konishi&Paffuti&Provero,Gross&Mende29} (examples of some other techniques can be found in \cite{Maggiore83,Maggiore82,Capozziello&Lambiase&Scarpetta,Ahluwalia31,Ahluwalia35}). There are other approaches to quantum gravity like the recently proposed Doubly Special Relativity (DSR) theories which suggest the presence of maximum observable momenta \cite{Amelino,Magueijo&Smolin03,Magueijo&Smolin10,Cortes&Gamboa}, connecting to minimum positions. Other branches of high energy physics such as the very early universe, or strong gravitational fields in black hole physics are also concerned about the minimal length \cite{Maggiore65}.
In fact, the usual Heisenberg Uncertainty Principle (HUP) fails for energies near the Planck scale, when the Schwarzschild radius is comparable to the Compton wavelength and both are close to the Planck length. This problem is resolved by revising the characteristic scale through the modification of HUP to what is known as the Generalized Uncertainty Principle (GUP) \cite{Kempf&Mangano&Mann,Kempf&Mangano}.
Among all complicated footprints of GUP, the most elegant description follows from the simple deductions of Newtonian and quantum gravity \cite{Adler&Santiago}, by considering a quantum particle such as electron, to be observed by photon in a thought instrument like the Heisenberg microscope. This elegancy explains why all of the arguments such as gedanken string collisions \cite{Amati&Ciafaloni&Veneziano,Gross&Mende29}, the thought experiment of black holes \cite{Maggiore65,Scardigli}, de Sitter space \cite{Snyder}, the symmetry of massless particle \cite{Chagas} and wave packets \cite{Bang&Berger}, agree that GUP holds at all scales as \cite{Amati&Ciafaloni&Veneziano,Garay,Maggiore65}

\begin{equation}\label{Eq0}
\Delta x_{i}\Delta p_{i}\geq\frac{\hbar}{2}\left[ 1+\beta \left( (\Delta p)^{2}+\langle p\rangle ^{2} \right)+\beta^{\prime} \left( (\Delta p_{i})^{2}+\langle p_{i}\rangle ^{2} \right) \right], i=1,2,3 ;
\end{equation}
where $p^{2}=\sum_{j=1}^{D} p_{j}^{2}$, $D$ is dimension of space, $\beta \sim l_{pl}^{2}/2\hbar^{2}$, $l_{pl}$ is Planck Length and $\beta^{\prime}$ is a constant.

Motivated by the above arguments, in this paper we try to study the influences of GUP on a Friedmann-Robertson-Walker (FRW) model of Hartle-Hawking
universe. The application of Einstein's field equations to the system of universe always faces with the problem of initial conditions. The Big Bang singularity is such a well-known problem in the standard model of cosmology. However, one can remove this problem by presenting a physical realization for the philosophical concept of a universe with no beginning.
This presentation was firstly made by Hartle and Hawking \cite{Hartle&Hawking}, where they showed that in the quantum interpretation of the very early
universe, it is not possible to express quantum amplitudes by 4-manifolds with globally Lorentzian geometries, instead they should be Euclidean compact manifolds with boundaries just located at a signature-changing hypersurface understood as the beginning of our Lorentzian universe.
This is well known as the {\it no boundary proposal}. In this direction of thinking about quantum interpretation of the early universe, many works have
also been accomplished on different cosmological models to study whether it is possible to realize a classical signature change  \cite{Dereli&Tucker,Darabi&Sepangi,Darabi,Darabi&Rastkar,Ghafoori&Gusheh&Sepangi} or not. Some of them have also considered the quantization of their models \cite{Darabi&Sepangi,Darabi,Darabi&Rastkar,Dereli&Onder&Tucker,Vakili&Jalalzadeh&Sepangi,Ahmadi&Jalalzadeh&Sepangi}. 
In a recent work \cite{Ghaneh&Darabi&Motavalli}, the special attention has
been paid for the case where the phase space coordinates are noncommutaive via the {\it Moyal product} approach. In the present work, we aim to study the effects of noncommutativity through the {GUP} approach in the phase space of a cosmological model which exhibits the signature change at the classical and quantum levels in the commutative case. We start with a FRW type metric and use a scalar field as the matter source of Einstein's field equations. Then, we apply the noncommutativity to the minisuperspace of corresponding effective action by the use of GUP approach in deforming the Poisson bracket. The conditions for which the classical signature change is possible are then investigated. Also, we study the quantum cosmology of this noncommutative signature changing model and find the perturbative solutions of the corresponding Wheeler-DeWitt equation. Finally, we investigate the interesting issue of classical-quantum correspondence in this model.

\section{Classical Signature Dynamics }

We consider a model of universe with the metric \cite{Dereli&Tucker}
\begin{equation}\label{Eq1}
g=-\varpi\, d\varpi \otimes d\varpi + \frac{\overline{R}^{2}(\varpi)}{1+(k/4)r^{2}}\,(dx^{i} \otimes dx^{i}),
\end{equation}
where $\overline{R}(\varpi)$ is the scale factor, $k=-1,0,1$ determines the spatial curvature. The sign of $\varpi$ is responsible for the geometry to be Lorentzian or Euclidian and the hypersurface of signature change is identified by $\varpi=0$. The cosmic time $t$ is related to $\varpi$ via $t=\frac{2}{3}\varpi^{3/2}$ when $\varpi$ is definitely positive.
One common way to treat the signature change problem is to obtain the exact solutions in Lorentzian region $(\varpi>0)$ and extrapolate them in Euclidian region continuously. In Lorentzian region, the line element (\ref{Eq1}) takes the form
\begin{equation}\label{Eq2}
ds^{2}=-dt^{2}+R^{2}(t)(dr^{2}+r^{2}d\Omega^{2}),
\end{equation}
where $k=0$ is set in agreement with the current observations. We also assume an scalar field with interacting potential $U(\phi)$ as the matter source. The corresponding action
\begin{equation}
{\cal S}=\frac{1}{2\kappa^{2}}\int d^{4}x\sqrt{-g}{\cal R}+\int d^{4}x\sqrt{-g}\left[-\frac{1}{2}(\nabla\phi)^{2}-U(\phi)\right]+{\cal S}_{YGH},
\end{equation}
leads to the following point like Lagrangian
\begin{equation}
{\cal L}=-3R\dot{R}^{2}+R^{3}\left[\frac{1}{2}\dot{\phi}^{2}-U(\phi)\right],
\end{equation}
where the units are adopted so that $\kappa\equiv1$ and the York-Gibbons-Hawking boundary term ${\cal S}_{YGH}$ is canceled by the surface terms \footnote
{Note that a dot determines differentiation with respect to $t$.}. A change of dynamical variables defined by
\begin{eqnarray}\label{6}
x_{1}=R^{3/2}cosh(\alpha\phi),
\end{eqnarray}
\begin{eqnarray}\label{7}
x_{2}=R^{3/2}sinh(\alpha\phi),
\end{eqnarray}
$(0\leq R<\infty,-\infty<\phi<+\infty)$ casts the Lagrangian into a more convenient form
\begin{equation}
{\cal L}=\dot{x}_{1}^{2}-\dot{x}_{2}^{2}+ 2\alpha^{2}U(\phi)(x_{1}^{2}-x_{2}^{2}),
\end{equation}
where $\alpha^{2}=\frac{3}{8}$, and a coefficient ``$-2\alpha^{2}$" is ignored by using the zero energy condition\footnote{According to Dirac's theory of
Hamiltonian constraint systems, general relativity is a constraint system whose constraint is the zero energy condition $H=0$ \cite{Dirac,ADM}.}.
Now, we choose the potential $U(\phi)$ \cite{Dereli&Tucker}
\begin{equation}\label{9}
2\alpha^{2}(x_{1}^{2}-x_{2}^{2})U(\phi)=a_{1}x_{1}^{2}+ a_{2}x_{2}^{2}+ 2b\,x_{1}x_{2},
\end{equation}
in which $a_{1},a_{2}$ and $b$ are constant parameters. Using (\ref{6}) and (\ref{7}), the potential is expressed in terms of $\phi$ 
\begin{equation}\label{10}
U(\phi)=\lambda+\frac{1}{2\alpha^2}m^2 \sinh^2(\alpha \phi)+\frac{1}{2\alpha^2}b \sinh(2\alpha \phi),
\end{equation}
where the physical parameters
\begin{eqnarray}\label{11}
&&\lambda=U\mid_{\phi=0}\,=a_{1}/2\alpha^{2},\\
&&m^{2}=\partial^{2}U/\partial\phi^{2}\mid_{\phi=0}\,=a_{1}+a_{2},
\end{eqnarray}
are defined as the cosmological constant and the mass of scalar field, respectively.
The Hamiltonian of system becomes
\begin{equation}\label{12}
{\cal H}(x,p)=\frac{1}{4}(p_{1}^{2}-p_{2}^{2})-a_{1}x_{1}^{2}-a_{2}x_{2}^{2}-2b\,x_{1}x_{2},
\end{equation}
where $p_{1}, p_{2}$ are the momenta conjugate to $x_{1},x_{2}$, respectively.
The dynamical equations $\dot{x}_{i}=\{x_{i},{\cal H}\}, (i=1,2)$ are then
written as \cite{Dereli&Tucker}
\begin{equation}
\ddot{\xi}={\sf M}\xi,
\end{equation}
where
\begin{equation}
{\sf M}=\left(
   \begin{array}{cc}
    a_{1} & b \\
    -b & -a_{2} \\
   \end{array}
  \right),\hspace{25mm}
\xi=\left(
        \begin{array}{c}
          x_{1} \\
          x_{2} \\
        \end{array}
      \right).
\end{equation}
In the normal mode basis ${\sf V}={\sf S}^{-1}\xi=\left( \begin{array}{c} {\sf q}_{1}\\ {\sf q}_{2}\\ \end{array} \right)$ for diagonalization of ${\sf M}$ as ${\sf S}^{-1}{\sf M}{\sf S}={\sf D}=diag({\sf m_{+}},{\sf m_{-}})$ we
find
\begin{equation}\label{Eq16'}
{\sf m_{\pm}}=\frac{3\lambda}{4}-\frac{m^{2}}{2}\pm\frac{1}{2}\sqrt{m^{4}-4b^{2}},
\end{equation}
and the solutions under initial conditions $\dot{{\sf V}}(0)=0$ are found
as
\begin{equation*}
{\sf q}_{1}(t)=2{\sf A}_{1}\cosh(\sqrt{{\sf m_{+}}}\,t),
\end{equation*}
\begin{equation}
{\sf q}_{2}(t)=2{\sf A}_{2}\cosh(\sqrt{{\sf m_{-}}}\,t),
\end{equation}
where ${\sf A}_{1},{\sf A}_{2}\in\mathbb{R}$. These solutions remain real when the phase of $(\sqrt{{\sf m_{+}}}\,t)$ changes by $\pi/2$, so they
are good candidates for real signature changing geometries. Note that the constants ${\sf A}_{1}$ and ${\sf A}_{2}$ are correlated by the zero energy condition
\cite{Dereli&Tucker}
\begin{equation}\label{V}
V^T(0)\mathcal I V(0)=0,
\end{equation}
where $\mathcal I={\sf S}^T {\sf J}{\sf M}{\sf S}$ and
$$ {{\sf J}=\left(\begin{array}{cc}
    1 & 0 \\
    0 & -1 \\
   \end{array}
  \right).}
$$
The equation (\ref{V}) is quadratic for the ratio $\chi={\sf A}_{1}/{\sf A}_{2}$ and its roots $\chi_{\pm}$ are determined by the parameters of $\lambda, m^2,b$. By choosing ${\sf A}_{2} =1$, the solutions fall into two following
classes
\begin{equation}\label{}
\xi^{\pm}(t)={\sf S}{\sf V}^{\pm}(t),
\end{equation}
where
\begin{equation}\label{}
{\sf q}^{\pm}_1(t)=2{\sf A}_{1}^{\pm}\cosh(\sqrt{{\sf m_{+}}}\,t),
\end{equation}
and
\begin{equation}\label{}
{\sf q}^{\pm}_2(t)=2\cosh(\sqrt{{\sf m_{-}}}\,t).
\end{equation}
At last, the original variables $R$ and $\phi$ are recovered from $x_{1}$ and $x_{2}$ via (\ref{6}) and (\ref{7}) as
\begin{equation}\label{}
R(t)=(x_{1}^2-x_{2}^2)^{1/3},
\end{equation}
\begin{equation}\label{}
\phi(t)=\frac{1}{\alpha}\tanh^{-1}\left(\frac{x_{2}}{x_{1}}\right).
\end{equation}

We conclude that: i) for both eigenvalues of {\sf M} being positive, no signature transition occurs, ii) for the product of the eigenvalues less than zero, the constraint (\ref{V}) is not satisfied with a real solution for the amplitude $\chi$, and iii) for both eigenvalues being negative, $x_{1}(\beta), x_{2}(\beta)$ exhibit bounded oscillations in the region $\beta> 0$ and are unbounded for $\beta< 0$ (see Fig.1 \cite{Dereli&Tucker}). Such behaviour is translated into the solutions for $R$ and $\phi$ (see Fig.2 \cite{Dereli&Tucker}). Therefore, it is possible to choose parameters so that the manifold becomes Euclidean for a finite range of $\beta< 0$ and undergoes a transition at $\beta= 0$ to become Lorentzian for a further finite range of $\beta> 0$ \cite{Dereli&Tucker}.

\section{Noncommutativity via deformation}

The study of noncommutativity between phase space variables is based on the
replacing of usual product between the variables with the star-product; and
in flat Euclidian spaces all the star-products are \emph{c-equivalent} to the so called Moyal product \cite{Hirshfeld}.\\
Let us assume $f(x_{1},..,x_{n};p_{1},..,p_{n})\,,g(x_{1},..,x_{n};p_{1},..,p_{n})$ to be two arbitrary functions. Then, the Moyal product is defined as
\begin{equation}\label{Eq17'}
f\star_{\propto}g=f\,e^{\frac{1}{2}\overleftarrow{\partial}_{a}\propto_{ab}\overrightarrow{\partial}_{b}}g,
\end{equation}
such that
\begin{equation}
\propto_{ab}=\left(
               \begin{array}{cc}
                 \theta_{ij} & \delta_{ij}+\sigma_{ij} \\
                 -\delta_{ij}-\sigma_{ij} &\bar{\theta}_{ij} \\
               \end{array}
             \right),
\end{equation}
and $\theta_{ij},\bar{\theta}_{ij}$ are antisymmetric $N\times N$ matrices.
Then, the deformed Poisson brackets read as
\begin{equation}
\{f,g\}_{\propto}=f\star_{\propto}g-g\star_{\propto}f.
\end{equation}
Therefore, the coordinates of a phase space equipped with Moyal product satisfy
\begin{equation}\label{Eq20}
\{x_{i},x_{j}\}_{\propto}=\theta_{ij},\hspace{20mm} \{x_{i},p_{j}\}_{\propto}=\delta_{ij}+\sigma_{ij},\hspace{20mm} \{p_{i},p_{j}\}_{\propto}=\bar{\theta}_{ij}.
\end{equation}
Considering the following transformations \cite{Masud}
\begin{equation}\label{Eq21}
x^{\prime}_{i}=x_{i}-\frac{1}{2}\theta_{ij}p^{j},\hspace{20mm} p^{\prime}_{i}=p_{i}+\frac{1}{2}\bar{\theta}_{ij}x^{j},
\end{equation}
one finds that $(x^{\prime}_{i},p^{\prime}_{j})$ fulfill the same commutation relations as (\ref{Eq20}) with respect to the usual Poisson brackets
\begin{equation}
\{x^{\prime}_{i},x^{\prime}_{j}\}=\theta_{ij},\hspace{20mm} \{x^{\prime}_{i},p^{\prime}_{j}\}=\delta_{ij}+\sigma_{ij},\hspace{20mm} \{p^{\prime}_{i},p^{\prime}_{j}\}=\bar{\theta}_{ij},
\end{equation}
provided that $(x_{i},p_{j})$ follows the usual commutation relations
\begin{equation}
\{x_{i},x_{j}\}=0,\hspace{20mm} \{p_{i},p_{j}\}=0,\hspace{20mm} \{x_{i},p_{j}\}=\delta_{ij}.
\end{equation}
This approach is so called \emph{noncommutativity via deformation}.

\section{Phase Space Deformation via GUP}

In this section, we aim to study the effects of noncommutativity in the phase space via deformation by GUP approach. The equation (\ref{Eq0}) represents a modification of Heisenberg algebra as
\begin{equation}\label{Eq15}
\left[ x^{\prime}_{i},p^{\prime}_{j} \right] = i\hbar \left(\delta_{ij}(1+\beta p^{\prime\,2})+\beta^{\prime}p^{\prime}_{i}p^{\prime}_{j} \right),
\end{equation}
where $\beta$,$\beta^{\prime}$ are taken to be small up to the first order. Then the ansatz of classical-quantum correspondence, $[ \,\,  ,  \, ] \rightarrow i\hbar \{ \,  , \, \}$, introduces the deformed poisson bracket of position coordinates and momenta \cite{NamChang&et.al.28} 

\begin{equation}\label{Eq16}
\{ x^{\prime}_{i},p^{\prime}_{j} \} = \delta_{ij}(1+\beta p^{\prime\,2})+\beta^{\prime}p^{\prime}_{i}p^{\prime}_{j},
\end{equation}
where primes on $x, p$ denotes the modified coordinates.
Assuming $\{p^{\prime}_{i},p^{\prime}_{j}\}=0$, the Jacobi identity almost uniquely specifies that \cite{Kempf&Mangano&Mann,Kempf}

\begin{equation}\label{Eq17}
\{ x^{\prime}_{i},x^{\prime}_{j} \} = \frac{(2\beta-\beta^{\prime})+(2\beta+\beta^{\prime})\beta p^{\prime\,2}}{1+\beta p^{\prime\,2}}(p^{\prime}_{i}x^{\prime}_{j}-p^{\prime}_{j}x^{\prime}_{i}).
\end{equation}
Remembering the usual (non-modified) algebra $\{x_{i},p_{j}\}=\delta_{ij}$, the relations (\ref{Eq16})-(\ref{Eq17}) can be realized by considering the following transformations

\begin{equation}\label{Eq19}
 x^{\prime}_{i}=(1+\beta p^{2})x_{i}+\beta^{\prime}p_{i}p_{j}x_{i}+\gamma\,p_{i}, \hspace{10mm} p^{\prime}_{i}=p_{i} .
\end{equation}
$\gamma$ being an arbitrary constant given by $\gamma=\beta+\beta^{\prime}\left(\frac{D+1}{2}\right)$
\cite{NamChang&et.al.27} .

\section{Signature Change in Deformed Phase Space}

 Let us follow the 2-dimensional model explained initially in section 2. The Hamiltonian of the deformed system is
\begin{equation}\label{Eq22}
{\cal H}^{\prime}(x^{\prime},p^{\prime}) =\frac{1}{4}(p^{\prime\,2}_{1}-p^{\prime\,2}_{2})-a_{1}x^{\prime\,2}_{1}-a_{2}x^{\prime\,2}_{2}-2bx^{\prime\,1}x^{\prime\,2},
\end{equation}

It can be described in terms of commutative coordinates by the use the transformations (\ref{Eq19}) as

\begin{equation}\label{Eq24}
{\cal H}^{\prime}(x,p)= {\cal W}(p)-{\cal Z}(p)^{2}\,{\cal U}(x)-2\gamma\,{\cal Z}(p){\cal V}(x,p),
\end{equation}
where $x_{i}, p_{j}$ reads the common Poisson algebra, and
\begin{eqnarray}
\nonumber && {\cal W}(p)=\frac{1}{4}\left[\left(1-4a_{1}\gamma^{2}\right)p_{1}^{2}-\left(1+4a_{2}\gamma^{2}\right)p_{2}^{2}\right]  , \\
\nonumber && {\cal U}(x)=a_{1}x_{1}^2+a_{2}x_{2}^2+2b x_{1}x_{2}  , \\
\nonumber && {\cal V}(x,p)=a_{1}x_{1}p_{1}+a_{2}x_{2}p_{2}+2b(x_{1}p_{2}+x_{2}p_{1})  , \\
&& {\cal Z}=1+\beta(p_{1}^{2}+p_{2}^{2})+\beta^{\prime}p_{1}p_{2}  .
\end{eqnarray}
It is usual to set $\beta^{\prime}=2\beta$ \cite{Vakili&Sepangi,Das,Vakili,Sepangi&Shakerin&Vakili} to make the shape of ${\cal Z}(p)$ more refined as ${\cal Z}({\cal P})$, ${\cal P}:=p_{1}+p_{2}$.

As is shown for a non-deformed system \cite{Dereli&Tucker} or the system deformed by \emph{moyal product} approach \cite{Ghaneh&Darabi&Motavalli}, the existence of a non-zero cross-term parameter $b$ in $U(\phi)$ is the only way to break the symmetry of the system under $\phi \rightarrow -\phi$
and make the change of signature happen.
However, we show that in contrary to the \emph{moyal product} approach, in {GUP} approach $b$ is not the only parameter responsible for signature change. To this end, we explicitly set $b=0$. On the other hand, to show that for
a continuous signature transition we need not choose a massive scalar field
we take a massless scalar field (i.e $a_{2}=-a_{1}$). By this set up we are
going to assert that a very specific scalar field potential of the form (\ref{10}) is not needed for a continuous signature transition. This makes continuous signature transition much easier than the model introduced in \cite{Dereli&Tucker}
because the justification of the complicate potential (\ref{10}) at early
universe is not a simple task. In the present model, however, we just need the
elements i) a free massless scalar field, ii) a cosmological constant,
and iii) GUP which are supposed to be trivial in the conditions at early universe.

The classical equations of motion $\dot{x}_{i}=\{x_{i},{\cal H}^{\prime}\}$, $i=1,2$, are then obtained as

\begin{eqnarray}\label{Eq26}
\nonumber && \dot{x_{1}}=4\beta\left(p_{1}+p_{2}\right)\left[{\cal Z}(p){\cal U}(x)+\gamma\,{\cal V}(x,p)\right] +2\gamma\, a_{1}x_{1}{\cal Z}(p)-\frac{1}{2}\left(1-4\gamma^{2}a_{1}\right)p_{1}, \\
          && \dot{x_{2}}=4\beta\left(p_{1}+p_{2}\right)\left[{\cal Z}(p){\cal U}(x)+\gamma\,{\cal V}(x,p)\right] -2\gamma\, a_{1}x_{2}{\cal Z}(p)+\frac{1}{2}\left(1-4\gamma^{2}a_{1}\right)p_{2} .
\end{eqnarray}
Also, the dynamical equations of momenta, $\dot{p}_{i}=\{p_{i},{\cal H}^{\prime}\}$, yield
\begin{eqnarray}\label{Eq27}
\nonumber && \dot{p_{1}}=-2a_{1}{\cal Z}(p)\left[x_{1}{\cal Z}(p)+\gamma\, p_{1}\right], \\
          && \dot{p_{2}}=\hspace{3mm} 2a_{1}{\cal Z}(p)\left[x_{2}{\cal Z}(p)+\gamma\, p_{2}\right] .
\end{eqnarray}
where a dot denotes differentiation with respect to $t$.

To decouple these equations, we merge (\ref{Eq26}) with (\ref{Eq27}) first, and then compute the summation and subtraction of the results. This procedure leads to the following equations

\begin{eqnarray}\label{Eq29}
\nonumber &&  8\beta_{1}^{2}\left(7{\cal Z}-8\right){\cal P}^{3}\dot{{\cal P}}^{6}-2{\cal Z}\left(27{\cal Z}^{2}-50{\cal Z}+24\right)\ddot{{\cal P}}\dot{{\cal P}}^{4}+2{\cal Z}^{2}\left(5{\cal Z}-4\right){\cal P}\dot{\ddot{{\cal P}}}\dot{{\cal P}}^{3}-{\cal Z}^{3}{\cal P}^{2}\ddot{\ddot{{\cal P}}}\dot{{\cal P}}^{2} \\
&&  -a_{1}^{2}\left(5{\cal Z}-4\right){\cal Z}^{6}{\cal P}^{3}\dot{{\cal P}}^{2}+2{\cal Z}^{3}{\cal P}^{2}\dot{\ddot{{\cal P}}}\ddot{{\cal P}}\dot{{\cal P}}-{\cal Z}^{3}{\cal P}^{2}\ddot{{\cal P}}^{3}+a_{1}^{2}{\cal Z}^{7}{\cal P}^{4}\ddot{{\cal P}}=0,
\end{eqnarray}
\begin{equation}\label{Eq30}
 p_{1}=\frac{1}{32a_{1}\beta^{2}\dot{{\cal P}}{\cal P}{\cal Z}}\left[a_{1}\beta{\cal Z}{\cal P}^{2}(3{\cal P}-16\beta\dot{{\cal P}})-{\cal Z}\ddot{{\cal P}}+a_{1}{\cal P}(1+\beta^{3}{\cal P}^{6})+4\beta{\cal P}\dot{{\cal P}}^{2}\right],
\end{equation}
\begin{eqnarray}\label{Eq32}
\nonumber && x_{1}=-\frac{1}{2a_{1}{\cal Z}^{2}}\left(8a_{1}\beta{\cal Z}p_{1}+\dot{p_{1}}\right), \\
          && x_{2}=-\frac{1}{2a_{1}{\cal Z}^{2}}\left(8a_{1}\beta{\cal Z}p_{2}-\dot{p_{2}}\right).
\end{eqnarray}

Eq.(\ref{Eq29}) is a differential equation with linear symmetry and it can be solved by order reduction via it's symmetry generators. Then the particular solution is obtained as

\begin{equation}\label{Eq33}
{\it RootOf}\left(2\int^{{\cal P}}\!{\frac{C_{1}}{\sqrt{-C_{1}\left(-4a_{1}^{2}y^{4}+4\,C_{1}^{2}C_{2}y^{2}+4C_{1}^{2}C_{2}^{2}y^{4}+C_{1}^{2}\right)
}\left(1+\beta y^{2}\right)}}{dy}\,+t+C_{3}\right),
\end{equation}
or equivalently
\begin{equation}\label{Eq34}
{\it RootOf}\left\{\Pi\left(C_{1}\beta/2\mathcal{C}_{+}; \arcsin(\sqrt{-2\mathcal{C}_{+}/C_{1}}{\cal P}), \sqrt{\mathcal{C}_{-}/\mathcal{C}_{+}}\right)-C_{1}\sqrt{\mathcal{C}_{+}/2}(t+C_{3})\right\},
\end{equation}
where $\Pi(\nu;\vartheta,\kappa)$ is the incomplete elliptic integral of the third kind, $\mathcal{C_{\pm}}=C_{1}C_{2}\pm a_{1}$, and $C_{1}$,$C_{2}$,$C_{3}$ are constants to be detected by initial conditions.

One can check that any such particular solution still remains a solution of (\ref{Eq29}) if it is multiplied by a minus sign, or (and) if any of the transformations $t\rightarrow -t$  or (and) $t\rightarrow it$ is applied . A simplified result is obtained at the special case where $C_{1}C_{2}=a_{1}$ \begin{equation}\label{Eq35}
{\cal P}={\frac{\sqrt{-C_{1}}\left({{\rm e}^{-\left(t+C_{3}\right)\Delta}}+1\right)}{\sqrt{\beta_{1}\,C_{1}\,\left({{\rm e}^{-\left(t+C_{3}\right)\Delta}}-1\right)^{2}+16\,a_{1}\,{{\rm e}^{-\left(t+C_{3}\right)\Delta}}}}},
\end{equation}
where $\Delta=\sqrt{-4\,a_{1}+\beta_{1}\,C_{1}}$.

\begin{figure}
  \includegraphics[width=8.5cm]{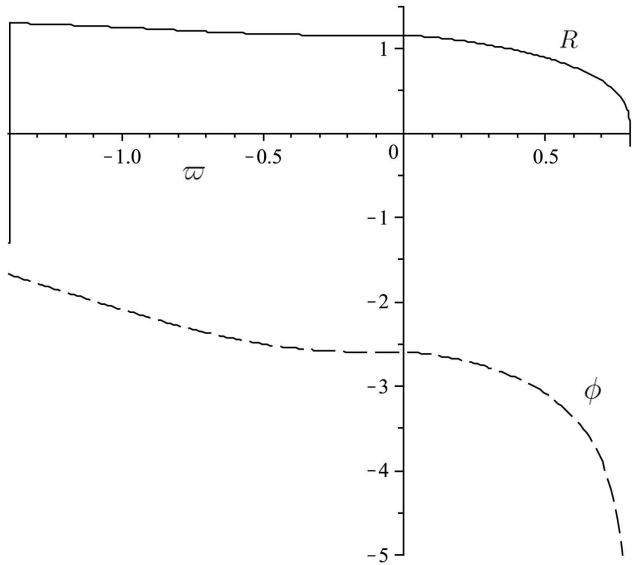}\\
  \caption{ {\small The real parts of scale factor (full curve) and scalar field (broken curve) in the \emph{first life} with respect to $\varpi$ for $\lambda=0.27, \beta=-0.45$.} } \label{Fig1}
\end{figure}

\begin{figure}
  \includegraphics[width=8.5cm]{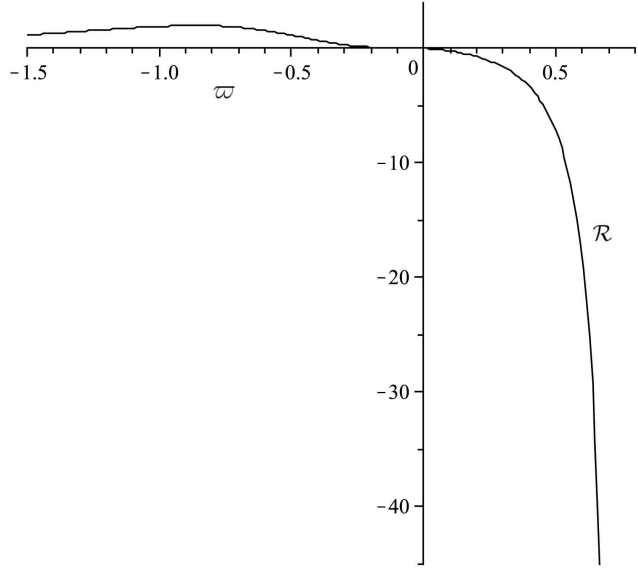}\\
  \caption{ {\small The behavior of Ricci scalar with respect to $\varpi$ for $\lambda=0.27, \beta=-0.45$.} } \label{Fig2}
\end{figure}

Physical values of $\lambda$ and $\beta$ ought to satisfy $\bar{{\cal R}}(0)=0$ and must also yield a positive $\bar{R}(\varpi)$ at the right neighborhood of $\varpi=0$, the area which can be called as \emph{Lorentzian region}.
The least requirement we expect is that the imaginary part of the physical functions $\bar{R}$,$\bar{\phi}$ and $\bar{\cal R}$ vanish at that area.
Fig.\ref{Fig1} and Fig.\ref{Fig2} show the signature transition by real solutions
from Euclidean to Lorentzian regions for a possible set of values\footnote{In these figures, the values of $\lambda, \theta$ and $C_{i}$ constants are finely selected in order to satisfy the mentioned requirements and the conditions ${\cal H}=0$ and ${\cal R}\mid_{_{\varpi=0}}=0$. We also note that changing the order of magnitude of these parameters does not affect the shape and physical behavior of these plots.}.

\section{Quantum Cosmology}

The high energy and small scale of very early universe provides the possibility
of having noncommutativity and GUP in the minisuperspace configurations of the model discussed here. But, in such small scale the quantum behavior is inevitable. Thus, it is necessary to study the quantized model and check if the quantization results are consistent with the classical solutions of dynamical equations.

Introducing the momentum quantum operators $\hat{p_{1}}=-i\partial/\partial x_{1},\,\hat{p_{2}}=-i\partial/\partial x_{2}$ and applying the Weyl symmetrization rule to (\ref{Eq24}) to construct the Hamiltonian operator, leads to the Wheeler-DeWitt(WD) equation of the form $\hat{{\cal H}}^{\prime}\Psi(x_{1},x_{2})=0$. Defining the real and imaginary parts of the wave function as $\Psi=\psi_{r}+i\psi_{i}$ splits WD equation in to two parts

\begin{equation}\label{Eq37}
H_{1}\psi_{r}-H_{2}\psi_{i}=0, \hspace{10mm} H_{2}\psi_{r}+H_{1}\psi_{i}=0,
\end{equation}
where,
\begin{eqnarray}\label{Eq38}
\nonumber&& H_{1}=8a_{1}\beta(x_{1}-x_{2})(\partial_{1}+\partial_{2})+a_{1}(x_{1}^{2}-x_{2}^{2})\left[2\beta(\partial_{1}+\partial_{2})^{2}-1\right]- \frac{1}{4}(\partial_{1}^{2}-\partial_{2}^{2}),\\
&& H_{2}=8a_{1}\beta(x_{1}\partial_{1}-x_{2}\partial_{2}).
\end{eqnarray}

In order to obtain a quantum criterion to test the classical results of previous section, we consider the special case $\psi_{r}={\cal A}\psi_{i}\equiv F(x_{1},x_{2})$, ${\cal A}$ being a constant. This converts (\ref{Eq37}) into $H_{1}F=0$
and $\, H_{2}F=0$, the second of which is automatically satisfied if $F=F(x_{1}x_{2})$, and the first one becomes

\begin{equation}\label{Eq39}
\left(2a_{1}\beta (x_{1}^{2}+X)^{2}+\frac{1}{4}x_{1}^{2}\right)\frac{d^{2}F}{dX^{2}}+12a_{1}\beta\,x_{1}^{2}\,\frac{dF}{dX}-a_{1}\,x_{1}^{2}\,F=0,
\end{equation}
where $X:=x_{1}x_{2}$ and $x_{1}$ is regarded as a parameter.
The solution of (\ref{Eq39}) is an expression of \emph{Generalized Hypergeometric Functions} as

\begin{eqnarray}\label{Eq40}
\nonumber &&F(x_{1},x_{2})= A_{1}\,\, _{2}\mathrm{F}_{1}\left(D_{+}(x_{1}),D_{-}(x_{1}) \,;\, -S^{\prime}(x_{1}) \,;\, \frac{1}{2}-a_{1}\beta S(x_{1})\left(1+\frac{x_{2}}{x_{1}}\right)\right) +  \\
\nonumber && A_{2}\,h(x_{1},x_{2})\,\,_{2}\mathrm{F}_{1}\left(S^{\prime}(x_{1})-D_{-}(x_{1}),S^{\prime}(x_{1})-D_{+}(x_{1}) \,;\, S^{\prime}(x_{1})+2 \,;\, \frac{1}{2}-a_{1}\beta S(x_{1})\left(1+\frac{x_{2}}{x_{1}}\right)\right),\\
&&
\end{eqnarray}
where $A_{1},A_{2}$ are two constants and

\begin{equation}\label{Eq41}
h(x_{1},x_{2}) = x_{1}^{2}\left(2\sqrt{-2a_{1}\beta}(x_{1}+x_{2})-1\right)^{S^{\prime}(x_{1})+S(x_{1})}.
\end{equation}

\begin{figure}
  \includegraphics[width=8.5cm]{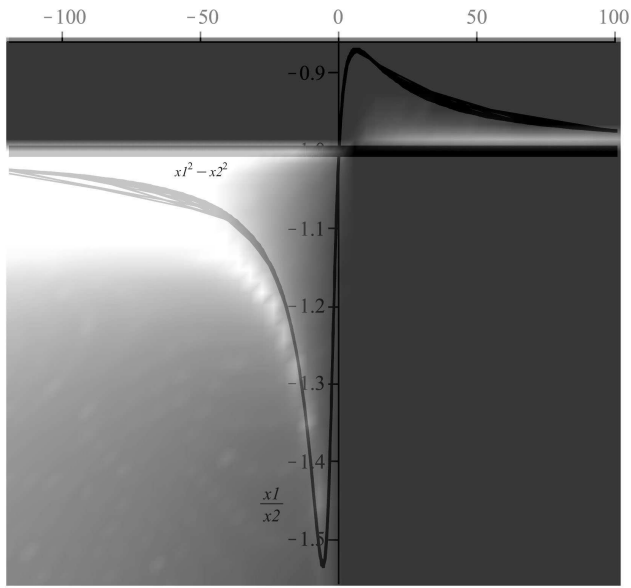}\\
  \caption{ {\small Density plot of $|\Psi|^{2}$ for $\lambda=0.27, \beta=-0.45$ which is in good agreement with the superimposed classical path.} } \label{Fig3}
\end{figure}

As figure \ref{Fig3} shows, the density plot of the quantum solution (\ref{Eq41}) is in good agreement with the classical solution obtained in the previous section.

\section{Conclusions}

Using a noncommutative phase space of dynamical variables deformed by Generalized Uncertainty Principle we have shown that continuous signature transition from Euclidean to Lorentzian may occurs for a model described by FRW metric minimally coupled with a free massless scalar field $\phi$ accompanied by a cosmological constant. The transformations of {GUP} in deforming the phase space breaks the symmetry of Hamiltonian under $\phi \rightarrow -\phi$ causing a possible continuous change of signature. This indicates that for
a signature transition to happen, instead of a massive scalar field having an {\it ad hoc} and complicate potential, we just need a free massless scalar field, a cosmological constant and a noncommutative phase space deformed
by GUP. These elements are supposed to be trivial in the extreme conditions at early universe. In commutative \cite{Dereli&Tucker} as well as \emph{moyal} transformed noncommutative Hamiltonian \cite{Ghaneh&Darabi&Motavalli}, we need a coupling $b$ in the scalar field potential to trigger the signature transition. However, using GUP in the absence of such potential and coupling, we have the expression $\beta^{\prime}p_{1}p_{2}$ in Hamiltonian (\ref{Eq24}) coming directly from the special structure of {GUP} deformations (\ref{Eq19}) which means that the {GUP} noncommutativity can cause a change of signature by itself. In other words, GUP accompanied by noncommutativity may establish
a general framework for a continuous change of signature. Moreover, in principle, the signature transition is possible for both negative and positive cosmological constants. This significantly differs from the \emph{moyal} approach \cite{Ghaneh&Darabi&Motavalli} in which only the negative values of cosmological constant are acceptable. We have also studied the quantum cosmology of this model and obtained a  solution of Wheeler-DeWitt equation showing a good correspondence with the classical path.

\end{document}